\documentclass[iop,preprint,showpacs,preprintnumbers,amsmath,amssymb]{revtex4}
\usepackage{graphicx}
\usepackage{dcolumn}
\usepackage{bm}

\begin{document}

\title{Tetragonal and trigonal deformations  in zinc-blende  semiconductors :
a tight-binding point of view}

\author{J.-M. Jancu$^{1}$ and P. Voisin$^{1}$}

\affiliation{$^1$ Laboratoire de Photonique et de Nanostructure, CNRS,
  route de Nozay, F-91000 Marcoussis, France}

\date{\today}

\begin{abstract}
 The deformation potentials of cubic semiconductors are re-examined  
from the point of view of the extended-basis  $sp^3d^5s^*$ tight-binding  model. 
Previous parametrizations had failed to account properly for trigonal deformations, 
even leading to incorrect sign of the acoustic component of the shear deformation potential {\it d}.
The strain-induced shifts and splittings of the on-site energies of
 the p- and d-orbitals are shown to play a prominent
 role in obtaining satisfactory values of deformation potentials both at the zone center and zone extrema.
 The present approach results in excellent agreement with
 available experimental data and ab-initio calculations.
\end{abstract}

\pacs{71.15.Ap, 71.55.Eq, 73.21.La, 85.75.-d}

\maketitle

The effect of uniaxial stress on the band structure of semiconductors 
has been a major theoretical and experimental topic for many years. 
With the development of strained-layer epitaxy, it has also become 
an important issue in modern material science and device physics. 
In their seminal approach, nearly half a century ago, Bir and Pikus established the strain Hamiltonian using the theory of invariants \cite{bir}. It depends on a number of deformation potentials describing the shifts and splittings of the various band extrema. For instance, for a given band  near the Brillouin zone center, it reads as :

\begin{eqnarray}
H_{\epsilon}^i =
-{\it a^i}\left( \epsilon_{xx}+\epsilon_{yy}+\epsilon_{zz}\right)-3{\it b^i}\left[\left({\it L_{z}^2} 
-\frac{1}{3}{\bf L}^2\right)\epsilon_{zz} +c.p \right]
-\frac{6 }{\sqrt 3}{\it d^i}\left[\left\{{\it L_{x}} {\it L_{y}}\right\}\epsilon_{xy}+c.p\right]
\end{eqnarray}

Where $\epsilon_{ij}$ are the components of the strain tensor, {\bf L} is the angular momentum operator, 
$\left\{{\it L_{x}} {\it L_{y}}\right\}=\frac{1 }{2}(\it L_{y}\it L_{x}+\it L_{x}\it L_{y})$, and c.p refers to circular permutations with respect to the axes $x$, $y$, $z$. The coefficient ${\it a^i}$ is the hydrostatic deformation potential for the ith band, while ${\it b^i}$ and  ${\it d^i}$ are respectively the tetragonal and rhombohedral (or trigonal) deformation potentials. We now explicit Eq. (1) for the $\Gamma_6$ and  $\Gamma_8$ states of zinc-blende crystals : the $\Gamma_6$ conduction-band energy only depends on the  hydrostatic term  owing of the ${\it L}=0$ matrix representation of the momentum operator. For the $\Gamma_8$ valence-band edge, the heavy- and light-hole degeneracy is lifted (as ${\it L}=1$) and the splitting depends on the strain orientation. Under [001] unixial stress or for lattice mismatched epilayers grown along the [001] direction, strain components can be written as:
$\epsilon_{xx}=\epsilon_{yy}$  $\neq$ $\epsilon_{zz}$  and 
$\epsilon_{xy}=\epsilon_{yz}=\epsilon_{zx}=0$. 
Thus, the heavy- and light-hole bands split by an amount proportional to ${\it b}$. 
Under [111] stress or 
for mismatched epilayers grown along the [111] direction, 
we have  $\epsilon_{xx}=\epsilon_{yy}=\epsilon_{zz}$$\ne0$ and 
$\epsilon_{xy}=\epsilon_{yz}=\epsilon_{zx}$$\ne0$, giving a valence-band splitting 
proportionnal to ${\it d}$. The two situations also differ by the presence or not of a static displacement of the anion and cation sublattices: in the former case all the atomic bonds in the strained crystal retain the same length and same angle
with respect to the strain symmetry axis, giving no short range contribution to the strain hamiltonian. 
Conversely, for trigonal distortions, the equilibrium positions of atoms are no longer fully determined 
by stress invariants and a relative displacement of the sublattices is allowed. 
The resulting internal strain is represented by the Kleinmann parameter 
$\zeta$ which ranges between 0 and 1 \cite{klei}. 
The value $\zeta=1$ corresponds a deformation with the same symmetry as the [111] strain but 
maintaining equal bond lengths of $a_0$${\sqrt 3}/4$, whereas $\zeta=0$ is related to the macroscopic strain 
that does not account for sublattice displacement. This gives simultaneously a 
long range or acoustic $\left({\it d'}\right)$ 
and a short range or optical $\left({\it d_{0}}\right)$ contribution to the  
rhombohedral deformation potential ${\it d}$  \cite{blacha}:

\begin{eqnarray}{\it d}= {\it d'}-\frac{\zeta}{4}{\it d_{0}} 
\end{eqnarray}

The strain hamiltonian has been extensively used in connection with the {\bf k.p} theory to 
interpret experimental data and measure parameters.  
On a more fundamendal side, several {\it ab initio} calculations  of the deformation potentials 
have been reported. More recently, atomistic approaches  using empirical parameters have 
become a strong challenger to k.p theory for a precise modeling of semiconductor nanostructures  
where compositions and deformations can vary rapidly at the bond-length scale. 
A remarkable feature of atomistic theories (as opposed to the fundamentally perturbative 
character of the k.p theory) is their natural ability to treat the whole Brillouin zone. 
It follows that a  "good" atomistic model must give a proper account of general distortions 
not only in the vicinity of the fundamental gap, but also at the edges of the Brillouin zone: 
the effects of strain actually is a very stringent test of atomistic models. 
Here we examine  the effects of tetragonal and trigonal deformations 
from the point of view of the empirical tight binding (TB) theory.\par 
Within the tight-binding formalism,  strain effects are mainly determined by scaling the 
Slater-Koster two-center integrals \cite{sk} (or transfer integrals) 
with respect to bond-length alterations, while bond-angle distortions are automatically 
incorporated via the phase factors in the Slater-Koster matrix elements. 
This leaves a more than sufficient number of strain-dependent parameters to 
fit the deformation potentials at the Brillouin zone center. 
However, when trying to fit simultaneously the splitting of zone-edge conduction valleys, 
one encounters a difficulty. For the case of [001] uniaxial strain, 
it was shown in Ref. \cite{jancu98} that adding a term corresponding to a 
strain-induced splitting of the $d$-orbitals  on-site energies (one-center integrals)   
leads to a much better overall fit of the deformation potentials at $\Gamma$ and $X$. 
Indeed, since the Wannier functions of tight-binding models are Slater-type orbitals \cite{kohn}, 
the one-center integrals are expected to be sensitive to the environment of neighboring atoms. 
In principle, one should also introduce a [001] shear parameter of the on-site $p$ energies, 
and this can be generalized to all diagonal matrix elements in proportion to cubic 
and uniaxial distortions. However, in the atomic limit, the on-site properties 
should depend neither on strain, nor on chemical environment. 
This is nearly the case for the $s$ and $p$ valence states that display an excellent 
degree of transferability. Conversely, it is clear that the excited $s$ and $d$-like orbitals 
have a strong free-electron character and corresponding energies must depend on strain-induced effects. 
Finally, one should remember that the "rule of the game" of atomistic models 
is to limit the number of empirical parameters to the minimum required to account 
for symmetries and reproduce experimental (or {\it ab initio} ) band parameters within a 
given accuracy. For instance, the hydrostatic shift of on-site energies appeared 
to be useless parameters at the present level of model sophistication, 
as they are for a large part renormalized in the variation of transfer integrals 
with bond-length changes.\par  
To the best of our knowledge, the case of trigonal deformations has never been discussed 
in the framework of an advanced tight binding model. 
A first difficulty is the choice of a value for the internal-strain parameter 
$\zeta$: contrarily to {\it ab initio} calculations, the atomic positions are an input of the 
tight binding model, not a result of the calculation. $\zeta$ can be obtained theoretically 
either in {\it ab initio} calculations or from the fit of phonon dispersions, 
as first demonstrated by Nielsen and Martin \cite{nielsen}. $\zeta$  was 
precisely measured  by X-ray diffraction for Si and Ge. 
The common value $\zeta$ = 0.54 \cite{cousins} is in agreement with first-principle calculations 
\cite{nielsen}. However, for GaAs, the  most-cited experimental result 
$\zeta=0.76$ \cite{kou} is still controversial \cite{nielsen,cardona87,zg,cousins1} 
and differs significantly from the theoretical value $\zeta$ = 0.48 obtained by 
Nielsen and Martin \cite{nielsen}.
Note that the latter value gives an 
excellent representation of elastic constants and phonon frenquencies of GaAs and is corroborated by 
 recent x-ray measurements which give $\zeta$=0.55 \cite{cousins1}. 
The second difficulty is methodological since a fit of {\it d} comes out of a calculation, 
while there are two quantities to determine, {\it d'}  and {\it $d_0$}.
Here, {\it d'}  is obtained by running the code using the fit parameters and setting $\zeta$ =0, 
whereas a relative displacement of the anion and cation sublattices in absence of macroscopic strain, 
is used to calculate {\it $d_0$}. \par 
The need for introducing a new shear parameter is evidenced by the failure 
of simpler tight-binding models : in the minimal $sp^3$ basis, 
the TB Hamiltonian cannot describe both {\it b} and {it d} satisfactorily, 
as seen for the diamond structure where {\it b}, {\it d'} , and  {\it $d_0$} 
are linked by the analytical relations \cite{blacha}: 
\begin{eqnarray}
{\it d_{0}}=16{\it d'}=-\frac{16 }{\sqrt 3} {\it b}\end{eqnarray}
For Ge, fitting {\it b}=-1.88 eV ${\pm 0.12}$ \cite{liu} and considering $\zeta=0.54$,  
one obtain: {\it d}=-1.26 eV, 
in poor agreement with the experimental result: {\it d}=-5.0 ${\pm 0.5}$  eV \cite{lb}. 
The failure is mainly caused by  an erroneous positive 
deformation potential ${\it d'}$ (see Eq. 3) in sharp contrast with the sign calculated by  
the  self-consistent LMTO and {\it ab initio} pseudopotential 
approaches \cite{cardona87,zg} which show  
a strong redistribution of the valence-electron 
density induced by the acoustic deformation.   
Similar discrepancies 
are obtained numerically for the zinc-blende semiconductors and no significant 
changes appear within the $sp^3s^*d^5$ approach, until a splitting of on-site energies is introduced. 
The corresponding hamiltonian obviously depends on the strain direction. 
For a uniaxial stress along [001], the perturbation has the $\Gamma_{12}$ symmetry, 
and the crystal field splits the fivefold degenerate $d$ orbitals into two doublets and a singlet as :

\begin{eqnarray}
E_{d_{xz}} = E_{d_{yz}} = E_d\left(1-\delta_{001}(\epsilon_{zz} - \epsilon_{xx}) \right)
\nonumber
\\
E_{d_{x^2-y^2}} = E_{d_{x^2-y^2}} = E_d
\\
E_{d_{xy}}= E_d\left(1+2\delta_{001}(\epsilon_{zz} - \epsilon_{xx}) \right)
\nonumber
\end{eqnarray}

For a uniaxial strain along [111] the perturbation has the $\Gamma_{15}$ symmetry and 
also splits the five equivalent $d$ bands 
into two doublets and a singlet state. 
To handle this case, it is more convenient to rotate the coordinate system 
and choose the quantization axis $\bar{z}$ along the [111] direction.
To avoid confusions we name "(111) basis" 
the new ( $\bar{x}$,  $\bar{y}$,  $\bar{z}$)  basis.
 
The on-site $d$ energies now corresponds to 
the representations $A_1$ with $E_{d_{3\bar{z}^2-r^2}}$, $E_1$ 
with $E_{d_{\bar{x}\bar{z}}}$ and $E_{d_{\bar{y}\bar{z}}}$, and $E_2$ with
$E_{d_{\bar{x}^2-\bar{y}^2}}$ and $E_{d_{\bar{x}\bar{y}}}$ : 

\begin{eqnarray}
E_{d_{3\bar{z}^2-r^2}}=E_d\left(1+2\delta_{111}(\epsilon_{\bar{z}} - \epsilon_{\bar{x}}) \right)
\nonumber
\\
E_{d_{\bar{x}\bar{z}}}=E_{d_{\bar{y}\bar{z}}}=E_d\left(1-\delta_{111}(\epsilon_{\bar{z}} 
- \epsilon_{\bar{x}}) \right)
\\
E_{d_{\bar{x}^2-\bar{y}^2}}=E_{d_{\bar{x}\bar{y}}}=E_d
\nonumber
\\
\epsilon_{\bar{z}} - \epsilon_{\bar{x}}=\frac{8}{3}(1-\zeta)\epsilon_{xy}
\nonumber
\end{eqnarray}

Negative $\epsilon_{xy}$ corresponds to conventional compressive stress. 
$\delta_{001}$ and $\delta_{111}$ are shear parameters fitted to reproduce 
the tetragonal and trigonal deformation of the valence-band edge, 
respectively. Note that, since perturbations result from different modifications 
of the nearest neighbors positions, there is no reason for an exact geometrical relation 
linking $\delta_{001}$ and $\delta_{111}$. 
The scheme of level splittings 
for uniaxial compressions along [001] and [111] is shown in figure 1. 
Following this procedure,we demonstrate excellent agreement with experiment for {\it b} and {\it d}. 
In addition, the figures coming out for the acoustic deformation potential {\it d'} 
are consistent with self-consistent LMTO results  \cite{cardona87}. 
For instance, for Ge, 
using the parametrization of Ref. \cite{jancu98}, with $\delta_{001}$=0.54 and $\delta_{111}$=-1.5 
, 
we get: {\it b} =-1.9 eV, {\it d} =-4.6 eV,  and {\it d'} =-1.37 eV, 
in agreement with experiment: {\it b}=-1.88 ${\pm 0.12}$ eV \cite{liu}, 
{\it d} =-5.0 eV ${\pm 0.5}$ \cite{lb}, 
and the LMTO calculation:  {\it d'}= -1.3 eV  \cite{cardona87}.
 
Furthermore, as discussed in Ref. \cite {jancu98}, 
introduction of $\delta_{001}$ allows to obtain simultaneously a fit of 
{\it b} and of the conduction-band splitting of $X$ valleys under [001] stress. 
However, for [111] stress, when examining the shear deformation potential of the 
$L_1$ conduction extrema, using our fit value of  $\delta_{111}$, 
we find a very large and disapointing discrepancy: $D_1^{5c}$=12.5 eV, to be compared to the experimental value 
of 18.3 eV \cite{lb}. This witnesses that there is still a missing parameter to describe properly the effects 
of trigonal deformations. From a quantum chemistry point of view, this discrepancy can be 
understood from the fact that the $L$-conduction band minimum is dominated by $s$ and $p$ states, 
in contrast with wavefunctions at $X$ that have a strong $d$ character \cite{jancu98}. 
Therefore, the splitting of on-site $d$  energies 
cannot help much to fit the strain deformation potential at $L$. Instead, the splitting of on-site $p$ energies, 
which was up to now a non-necessary parameter, becomes important to model strain field anisotropies. At this point, 
the internal logic of the model implies that if the strain induced splittings of 
the one-center integrals play such an important role, their shift under hydrostatic strain should 
also be taken explicitely into account instead of being renormalized in a strain dependency of 
two-center integrals. 
Using again the "(111) basis", the corresponding contributions to (111)-strain hamiltonian are written as:

\begin{eqnarray}
E_{p_{\bar{x}}} = E_{p_{\bar{y}}} = E_p\left(1-\pi_{111}(\epsilon_{\bar{z}} - \epsilon_{\bar{x}}) \right)
\nonumber
\\
E_{p_{\bar{z}}}= E_p\left(1+2\pi_{111}(\epsilon_{\bar{z}} - \epsilon_{\bar{x}}) \right)
\end{eqnarray}

The values of ${\it d}$, ${\it d'}$, ${\it d_{0}}$, and $D_1^{5c}$,  for Ge and GaAs semiconductors are
compared in table 1 with available experimental data and 
LMTO \cite{cardona87} evaluations. 
We use values of $\zeta$ issued from experiments and fit
$\pi_{111}$ and $\delta_{111}$ to  reproduce
the trigonal deformation of the valence-band edge and $L$-conduction band valleys. 
To achieve a complete description 
of strain effects in the $spds*$ model, 
on-site Hamiltonian matrix elements were also scaled with respect to bond-length changes resulting.
Althgouh the number of tight-binding parameters is increased with respect of Ref \cite{jancu98} their numerical determination 
through  a multi-parameter fitting procedure still converges very well and resulting values gain a more intuitive view of
chemical dependences.
A detailed description of the complete  parametrization 
is beyond the scope of the present work and will be the subject of a forthcoming paper. 
The data presented in Table 1 demonstrate that our results are in good agreement with experimental and 
theorerical values for $d$ an $D_1^{5c}$. The magnitude and sign of ${\it d'}$ and ${\it d_{0}}$ 
are well reproduced compared to the LMTO calculations. Our values for ${\it d_{0}}$ differ from the experimental results 
whoses values are actually contreversial, a discrepancy observed already
earlier; (compare Table 1).\\

In conclusion, we have demonstrated that the $sp^3d^5s^*$ model requires diagonal matrix element shifts 
to correctly reproduce uniaxial [klm] strain for cubic semiconductors. 
In order to test our model we have calculated the acoustic and optical contributions 
to the trigonal deformation potentials and found a good agreement 
with experiment and LMTO results.
A major improvement compared to smaller TB models was the correct sign and magnitude of the 
acoustic deformation potential
${\it d'}$ directly  related to the shear parameter of $d$-states. This TB
model provides a valid framework for the calculation of strain effects in self-assembled quantum dots.

\acknowledgments
The authors thank F. Glas for clarifying discussions.

\newpage
\begin{figure}
\vspace{0.5cm}
\includegraphics[angle=0,width=8cm]{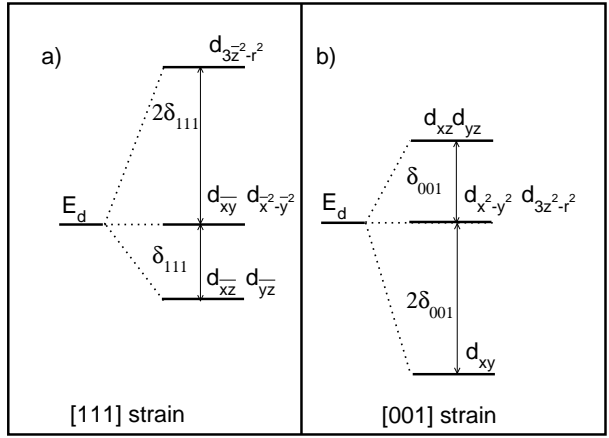}
\vspace{0.5cm}
\caption{Schematic plot of energy-level splitting of {\it d} states induced by 
an uniaxial strain along the [111] (left panel) and [001] directions (right panel). For [111] strain, 
the quantization axis $\bar{z}$ is chosen along the [111] direction.}
\end{figure}

\begin{table}
\caption{ Comparison of
  trigonal deformation potentials obtained in present work with LMTO 
 and experimental results. We use $\zeta=0.54$ for Ge and $\zeta=0.55$ for GaAs.}

\begin{tabular}{cccccccccccc} \\ \hline \hline
&&&Ge&&&&GaAs& \\ \hline
&$d_0$&$d'$&$d $&$D_1^{5c}$&$d_0$&$d'$&$d$&$D_1^{5c}$  \cr \hline
this work  & 30 & -0.9& -5.0& 17.1&27&-0.8&-4.5&18.0&      \\ 
LMTO $^a$ & 22.4&-1.3 &-5.0& & 25&-0.99&&&\\ 
Expt.$^b$&33& &-5.0&18.4&44&&-4.5&20.8 \\ 

\hline \hline

\end{tabular}

$^a$  Ref. \cite{cardona87},\
$^b$ Landolt-B \"{o}rnstein Ref. \cite{lb}.\\ 
\end{table}

\end{document}